\documentclass[aip,reprint,superscriptaddress,amsmath,amssymb,twocolumn,preprintnumbers]{revtex4-2}

\usepackage{graphics,graphicx}

\begin{document}
\preprint{Physics of Fluids}

\title{Momentum flux fluctuations in wall turbulence: a formula beyond the law of the wall}

\author{Hideaki Mouri}
\email[Corresponding author.]{}
\affiliation{Meteorological Research Institute, Nagamine, Tsukuba 305-0052, Japan}
\affiliation{Graduate School of Science, Kobe University, Rokkodai, Kobe 657-8501, Japan}
\author{Junshi Ito}
\affiliation{Meteorological Research Institute, Nagamine, Tsukuba 305-0052, Japan}
\affiliation{Graduate School of Science, Tohoku University, Aoba, Sendai 980-8578, Japan}



\begin{abstract}
Within wall turbulence, there is a sublayer where the mean wall-normal flux of the streamwise momentum is constant and related to the logarithmic wall-normal profile of the mean streamwise velocity. This relation, i.e., the law of the wall, has been used to estimate the mean stress at the wall surface. However, the momentum flux exhibits large temporal fluctuations. To relate them theoretically to those of the streamwise velocity at the same position from the wall, we consider an orthogonal decomposition of the fluctuations on a plane of the streamwise and wall-normal velocities. Since a large timescale is expected for the component that would dominate the momentum flux, it is singled out by temporal smoothing. The resultant formula is consistent with time-series data of a boundary layer in a wind tunnel. We also extend the formula to thermally stratified cases.
\end{abstract}

\maketitle

\section{Introduction} \label{S1}

Wall turbulence is a system to transfer the momentum of the flow toward the wall surface. If homogeneous over the wall and stationary, at least in some extents, the mean rate of that transfer or the mean momentum flux is related to the mean streamwise velocity via the law of the wall.\cite{my71} However, as for fluctuations of the momentum flux and the streamwise velocity, such a law is still uncertain. We study it theoretically and experimentally.

The configuration is as follows. We take the $x$-$y$ plane at the wall surface. The $x$ direction is that of the mean stream. For a given position $(x, y)$, while $U(z)$ denotes the mean velocity at a distance $z$ from the wall surface, $u(z, t)$, $v(z, t)$, and $w(z, t)$ denote fluctuating velocities at time $t$ in the streamwise, spanwise, and wall-normal directions. Also, $\delta$ is the thickness of the turbulence.

We assume an incompressible flow and study its momentum flux per unit mass density. Our focus is on $uw$, which corresponds to the rate of the wall-normal transfer of the streamwise momentum.

The surface of the wall is either smooth or rough. A rough surface is parameterized by the length $z_0$ of the aerodynamic roughness. For a smooth surface, the kinematic viscosity $\nu$ could be used to define $z_0$ as $\varpropto \nu/u_{\ast}$, where $u_{\ast}$ is the friction velocity.\cite{my71}

The turbulence is assumed to be stationary in the strict sense. We take an average always over time $t$. For example, $\langle X \rangle$ is the temporal average of a quantity $X(t)$.

Asymptotically at $z_0 \ll z \ll \delta$ in the limit $\delta/z_0 \rightarrow +\infty$,\cite{my71} there is a sublayer where the mean momentum flux $\langle uw \rangle$ is constant at the value of $-u_{\ast}^2 < 0$. Even if $\delta/z_0$ is finite, so long as it is large enough, such a constant-flux sublayer is still a good approximation for a range of distances $z$.

Throughout this constant-flux sublayer, the friction velocity $u_{\ast}$ serves as a single characteristic velocity. Since there is no constant in units of length, Landau\cite{ll59, my71} has pointed out a relation
\begin{subequations} \label{eq1}
\begin{equation} \label{eq1a}
\frac{z}{u_{\ast}} \frac{dU}{dz} = \frac{1}{\kappa} .
\end{equation}
For the von K\'arm\'an constant $\kappa$, which lies at $0.39 \pm 0.02$ in various configurations of wall turbulence, e.g., boundary layer, pipe flow, and channel flow,\cite{mmhs13} we assume the value of $0.40$. The integration of Eq.~(\ref{eq1a}) leads to
\begin{equation} \label{eq1b}
\frac{U(z)}{u_{\ast}} = \frac{1}{\kappa} \ln \left( \frac{z}{z_0} \right) ,
\end{equation}
or equivalently
\begin{equation} \label{eq1c}
\langle uw \rangle = -u_{\ast}^2 = - \frac{\kappa^2}{\ln^2 (z/z_0)} U^2(z)  .
\end{equation}
\end{subequations}
This is the law of the wall,\cite{my71} where the mean velocity $U$ is related to the mean momentum flux $\langle uw \rangle$. It is equal to the mean stress at the wall surface.

For instantaneous values of the momentum flux $uw$ and the streamwise velocity $U+u$, Deardorff\cite{d70} has proposed a model by analogy with Eq.~(\ref{eq1c}),
\begin{equation} \label{eq2}
uw(z, t) = - \frac{\kappa^2}{\ln^2 (z/z_0)} \left[ U(z)+u(z, t) \right]^2 .
\end{equation}
This model is often used to estimate the wall stress at each time $t$.\cite{d70, bmp05, kl12, lkbb16, ypm17, bp18, bl21, im21} However, it is too approximate. We need smoothing in space\cite{bmp05, ypm17, bp18} or in time\cite{ypm17, bp18} so as to have $uw$ and $U+u$ in Eq.~(\ref{eq2}) close to their averages $\langle uw \rangle$ and $U$ in Eq.~(\ref{eq1c}).

Since the fluctuations of the momentum flux $uw$ are crucial to our understanding of wall turbulence,\cite{my71, wl72, wbe72, lw73, wb77, amt00, fsc07, nkpk07, dn11b, w16, dm21, im21} we explore a formula beyond the law of the wall. The result is
\begin{subequations} \label{eq3}
\begin{equation} \label{eq3a}
\overline{uw}_{\tau}(z,t) - \langle uw \rangle = -\kappa \left[ \overline{u^2}_{\tau}(z,t) - \langle u^2(z) \rangle \right] ,
\end{equation}
with
\begin{equation} \label{eq3b}
\overline{X}_{\tau}(z,t) = \frac{1}{\tau} \int^{+\tau/2}_{-\tau/2} X(z, t+t') dt' .
\end{equation}
\end{subequations}
Here $\langle uw \rangle$ is to be obtained from Eq.~(\ref{eq1c}). For smoothing in Eq.~(\ref{eq3b}), a timescale $\tau \gtrsim 10 z/U$ is necessary\cite{im21} to remove fluctuations that have not been incorporated into Eq.~(\ref{eq3a}).

The derivation of Eq.~(\ref{eq3}) is described in Sec.~\ref{S2}. With use of experimental time-series data of a boundary layer (Sec.~\ref{S3}), we confirm Eq.~(\ref{eq3}) in Sec.~\ref{S4}. It is related to a phenomenology of energy-containing eddies\cite{t76, ka99, amt00, dn11a, dn11b, bhm17, mouri17, mmym17, mm19} and also extended to thermally stratified cases via the theory of Monin and Obukhov\cite{my71, f06} (Sec.~\ref{S5}). Finally, in Sec.~\ref{S6}, we conclude with remarks on applications of Eq.~(\ref{eq3}).

\section{Model} \label{S2}

\begin{figure}[tbp]
\begin{center}
\resizebox{8.6cm}{!}{\includegraphics*[3.3cm,20.2cm][18.0cm,27.0cm]{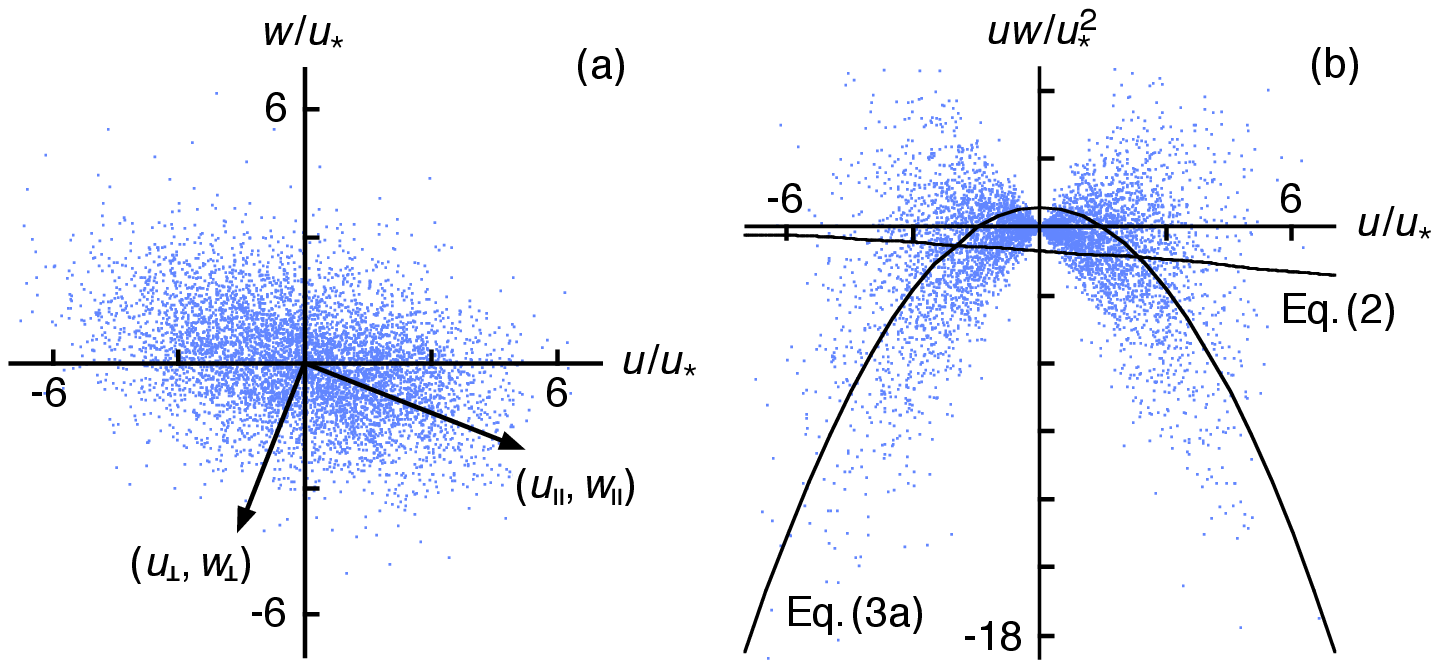}}
\caption{\label{f1} Schematics of Eq.~(\ref{eq4}) on the $u$-$w$ plane (a) and of Eqs.~(\ref{eq2}) and (\ref{eq3a}) for $\tau = 0$ on the $u$-$uw$ plane (b). Scattered points are from our data at $z/\delta_{99} = 0.15$.}
\end{center}
\end{figure}

For the constant-flux sublayer, instantaneous values of the velocity fluctuations $u$ and $w$ at a given distance $z$ are known to scatter elliptically on the $u$-$w$ plane with the major axis lying from the second quadrant $u<0$ and $w>0$ to the fourth quadrant $u>0$ and $w<0$.\cite{my71, wb77, w16, im21} To model such fluctuations, we decompose the vector $(u, w)$ into two orthogonal vectors $(u_{\shortparallel} , w_{\shortparallel})$ and $(u_{\perp} , w_{\perp})$ as
\begin{equation} \label{eq4}
\left.
\begin{array}{ll}
u =\!&u_{\shortparallel}+ u_{\perp} ,\\
w =\!&w_{\shortparallel}+ w_{\perp} = -\kappa u_{\shortparallel} + u_{\perp}/\kappa .
\end{array}
\right.
\end{equation}
These two vectors are aligned well with the major and minor axes of the above ellipse as in Fig.~\ref{f1}(a).

To the mean momentum flux $\langle uw \rangle <0$, the major component $u_{\shortparallel}$ with $w_{\shortparallel} = -\kappa u_{\shortparallel}$ is predominant. The inclination $w_{\shortparallel}/u_{\shortparallel} = -\kappa$ is from $dU/dz = u_{\ast}/\kappa z$ in Eq.~(\ref{eq1a}). If a fluid particle moves by a wall-normal distance ${\mit\Delta}z$, its typical contribution to the streamwise velocity is ${\mit\Delta}u =$ $-(dU/dz) {\mit\Delta}z$. The wall-normal velocity of that motion is ${\mit\Delta}z/(z/u_{\ast})$ because the characteristic timescale is $z/u_{\ast}$. From $u_{\shortparallel} = {\mit\Delta}u$ and $w_{\shortparallel} = {\mit\Delta}z/(z/u_{\ast})$, we have obtained $w_{\shortparallel}/u_{\shortparallel} = -\kappa$ (see also Sec.~\ref{S5a}).

The minor component $u_{\perp}$ with $w_{\perp} = u_{\perp}/\kappa$ is regarded here as random disturbances, i.e., $u_{\perp}w_{\perp} > 0$ against $\langle uw \rangle < 0$. Since the magnitudes of its fluctuations are not large, its timescale would not be large also. Actually, according to time-series analyses, fluctuations in the first quadrant $u>0$ and $w>0$ and the third quadrant $u<0$ and $w<0$ tend to have shorter durations than those in the other quadrants.\cite{wl72, wbe72, nkpk07, w16, dm21}

We use Eq.~(\ref{eq3b}) to smooth $uw - \langle uw \rangle$ over a timescale $\tau$. The result is written with $\overline{X}\vert_{\tau} = \overline{X}_{\tau}$ as
\begin{subequations} \label{eq5}
\begin{align} \label{eq5a}
\overline{uw}_{\tau} - &\langle uw \rangle = \nonumber \\
                       &\overline{u_{\shortparallel}w_{\shortparallel}}\vert_{\tau} - \langle u_{\shortparallel}w_{\shortparallel} \rangle  
                      + \overline{u_{\shortparallel}w_{\perp}         }\vert_{\tau} - \langle u_{\shortparallel}w_{\perp}          \rangle  \\
                      +&\overline{u_{\perp}         w_{\shortparallel}}\vert_{\tau} - \langle u_{\perp}         w_{\shortparallel} \rangle  
                      + \overline{u_{\perp}         w_{\perp}         }\vert_{\tau} - \langle u_{\perp}         w_{\perp}          \rangle  \nonumber. 
\end{align}
If the minor component $(u_{\perp}, w_{\perp})$ is independent of the major component $(u_{\shortparallel}, w_{\shortparallel})$ and if its fluctuations do have shorter durations, $\overline{u_{\shortparallel}w_{\perp}}\vert_{\tau}$, $\overline{u_{\perp}w_{\shortparallel}}\vert_{\tau}$, and $\overline{u_{\perp} w_{\perp}}\vert_{\tau}$ converge to their averages faster than $\overline{u_{\shortparallel}w_{\shortparallel}}\vert_{\tau}$. For a large enough timescale $\tau$,
\begin{equation} \label{eq5b}
\left.
\begin{array}{rlll}
\overline{u_{\shortparallel}w_{\perp}}\vert_{\tau}& \rightarrow  &\langle u_{\shortparallel}w_{\perp}          \rangle &=0,         \\
\overline{u_{\perp}w_{\shortparallel}}\vert_{\tau}& \rightarrow  &\langle u_{\perp}         w_{\shortparallel} \rangle &=0,         \\
\overline{u_{\perp}w_{\perp}}         \vert_{\tau}& \rightarrow  &\langle u_{\perp}         w_{\perp}          \rangle          
                                                  & =             \langle u_{\perp}^2                          \rangle /\kappa > 0, 
\end{array}
\right.
\end{equation}
but yet
\begin{equation} \label{eq5c}
\overline{u_{\shortparallel}w_{\shortparallel}}\vert_{\tau} \nrightarrow \langle u_{\shortparallel}w_{\shortparallel} \rangle
                                                            =    -\kappa \langle u_{\shortparallel}^2                 \rangle < 0, 
\end{equation}
and hence
\begin{equation} \label{eq5d}
\overline{uw}_{\tau} - \langle uw \rangle = \overline{u_{\shortparallel}w_{\shortparallel}}\vert_{\tau} - \langle u_{\shortparallel}w_{\shortparallel} \rangle .
\end{equation}
Thus $\overline{uw}_{\tau}$ is equal to $\langle uw \rangle - \langle u_{\shortparallel}w_{\shortparallel} \rangle + \overline{u_{\shortparallel}w_{\shortparallel}}\vert_{\tau}= \langle u_{\perp}w_{\perp}\rangle + \overline{u_{\shortparallel}w_{\shortparallel}}\vert_{\tau}$ with $\langle u_{\perp}w_{\perp}\rangle  > 0$ and $\overline{u_{\shortparallel}w_{\shortparallel}}\vert_{\tau} = -\kappa \overline{u_{\shortparallel}^2}\vert_{\tau} < 0$. We rewrite Eq.~(\ref{eq5d}) as
\begin{equation} \label{eq5e}
\overline{uw}_{\tau} - \langle uw \rangle = -\kappa \overline{u_{\shortparallel}^2}\vert_{\tau} + \kappa \langle u_{\shortparallel}^2 \rangle .
\end{equation}
\end{subequations}
Likewise,
\begin{equation} \label{eq6}
\overline{u^2}_{\tau} - \langle u^2 \rangle = \overline{u_{\shortparallel}^2}\vert_{\tau} - \langle u_{\shortparallel}^2 \rangle  .
\end{equation}
From Eqs.~(\ref{eq5e}) and (\ref{eq6}), we derive the final result between $\overline{uw}_{\tau} - \langle uw \rangle$ and $\overline{u^2}_{\tau} - \langle u^2 \rangle$ in the form of Eq.~(\ref{eq3a}).

This modeling does not apply to the flux $vw$ of the spanwise velocity $v$. From $\langle vw \rangle =0$, it is implied that $vw$ does not have a dominant component corresponding to $u_{\shortparallel}$ with $w_{\shortparallel} = - \kappa u_{\shortparallel}$ in Eq.~(\ref{eq4}).

Figure~\ref{f1}(b) compares Eq.~(\ref{eq3a}) with the existing model of Eq.~(\ref{eq2}). Throughout the present range of $u$, while $uw$ in Eq.~(\ref{eq2}) is decreasing with $u$ and is always negative, $\overline{uw}_{\tau}$ in Eq.~(\ref{eq3a}) is convex upward and becomes positive when $\langle u_{\perp}w_{\perp}\rangle > 0$ dominates over $\overline{u_{\shortparallel}w_{\shortparallel}}\vert_{\tau} < 0$. These two models are both inconsistent with actual data of instantaneous fluctuations $uw$ (scattered points). Nevertheless, with smoothed fluctuations $\overline{uw}_{\tau}$, Eq.~(\ref{eq3a}) is consistent as described in Sec.~\ref{S4}.

\section{Experimental Data} \label{S3}

The data studied here are those from our experiment of a boundary layer in a wind tunnel of the Meteorological Research Institute. We use coordinates $x_{\rm wt}$, $y_{\rm wt}$, and $z_{\rm wt}$ in the streamwise, spanwise, and floor-normal directions. The origin $x_{\rm wt} = y_{\rm wt} = z_{\rm wt} = 0$ is taken at the center of the floor at the upstream end of the measurement section of the wind tunnel. Its size is ${\mit\Delta}x_{\rm wt} = 22$\,m, ${\mit\Delta}y_{\rm wt} = 3$\,m, and ${\mit\Delta}z_{\rm wt} = 2$\,m. Upon the entire floor from $x_{\rm wt} = 0$ to $22$\,m with an interval of ${\mit\Delta}x_{\rm wt} = 50$\,mm, spanwise rods of diameter $2$\,mm were set as roughness.

The incoming wind velocity was $12$\,m\,s$^{-1}$. Downwind at $x_{\rm wt} =$ $18$\,m, the boundary layer was well developed and was almost homogeneous in the $x_{\rm wt}$ direction. There, at $y_{\rm wt} = 0$\,m and $z_{\rm wt} = 18$ to $600$\,mm, we obtained time series of the streamwise velocity $U+u$ simultaneously with the floor-normal velocity $w$.

\begin{table}[bp]
\begingroup
\squeezetable
\caption{\label{t1} Characteristics of the boundary layer. The uncertainties are $\pm 2 \sigma$ errors.}
\begin{ruledtabular}
\begin{tabular}{lcc}
Quantity                                              & Unit             & Value              \\  \hline
Turbulence thickness $\delta_{99}$                    & mm               & $373  \pm  3$      \\    
Friction velocity $u_*$                               & mm\,s$^{-1}$     & $542  \pm  1$      \\
Kinematic viscosity $\nu$                             & mm$^2$\,s$^{-1}$ & $15.3 \pm  0.0$    \\
Roughness length $z_0$ in Eq.~(\ref{eq1b})            & mm               & $0.10 \pm  0.02$   \\ 
von K\'arm\'an constant $\kappa$  in Eq.~(\ref{eq1b}) &                  & $0.41 \pm  0.01$   \\
$c_{u^2}$ in Eq.~(\ref{eq10})                         &                  & $2.43 \pm  0.19$   \\
$d_{u^2}$ in Eq.~(\ref{eq10})                         &                  & $1.17 \pm  0.06$   \\
$c_{w^2}$ in Eq.~(\ref{eq10})                         &                  & $1.61 \pm  0.05$   \\
\end{tabular}
\end{ruledtabular}
\endgroup
\end{table}

\subsection{Measurement and data processing} \label{S3a}

We used a hot-wire anemometer made up of a constant temperature circuit (Dantec, 90C10) and a crossed-wire probe (Dantec, 55P53). Its two wires were of platinum-plated tungsten, $5$\,$\mu$m in diameter, $1.25$\,mm in sensing length, $1$\,mm in separation, and at $\pm45^{\circ}$ to the streamwise direction. The overheat ratio was $0.95$. We calibrated the anemometer before and after each series of the measurements.

The anemometer signal was low-pass filtered at a cut-off frequency $f_{\rm c}$ and then sampled at a frequency $2f_{\rm c}$. We set $f_{\rm c}$ as high as possible at each distance $z_{\rm wt}$, provided that noise was not significant throughout the power spectrum.

At the distances $z_{\rm wt} = 32$ to $75$\,mm, which were expected to lie within or around the constant-flux sublayer, the total lengths of the individual time series were $24 \times$ $10^7$ in the number of the pairs $U+u$ and $w$. They were sampled at $2f_{\rm c} = 40$\,kHz. Those at the other distances were $6 \times 10^7$ at $2f_{\rm c} = 20$ to $44$ kHz.

Our analyses are on segments of the individual time series. Their lengths are $1 \times 10^7$. Among them, values of any statistic scatter because of incomplete convergence due to a limited sampling duration. We regard such segments as multiple independent realizations of the turbulence and use that scatter to calculate the final errors in a usual manner.\cite{br03}

To estimate the thickness of the boundary layer $\delta$ as a distance $z_{\rm wt} = \delta_{99}$ at which the mean velocity $U$ is $99$\% of its maximum,  we obtained short data of $U$ with an interval of ${\mit\Delta}z_{\rm wt} = 10\,\mbox{mm}$. Their errors are calculated as described above about the long data.

\begin{figure}[tbp]
\begin{center}
\resizebox{8.6cm}{!}{\includegraphics*[2.3cm,10.1cm][17.8cm,26.7cm]{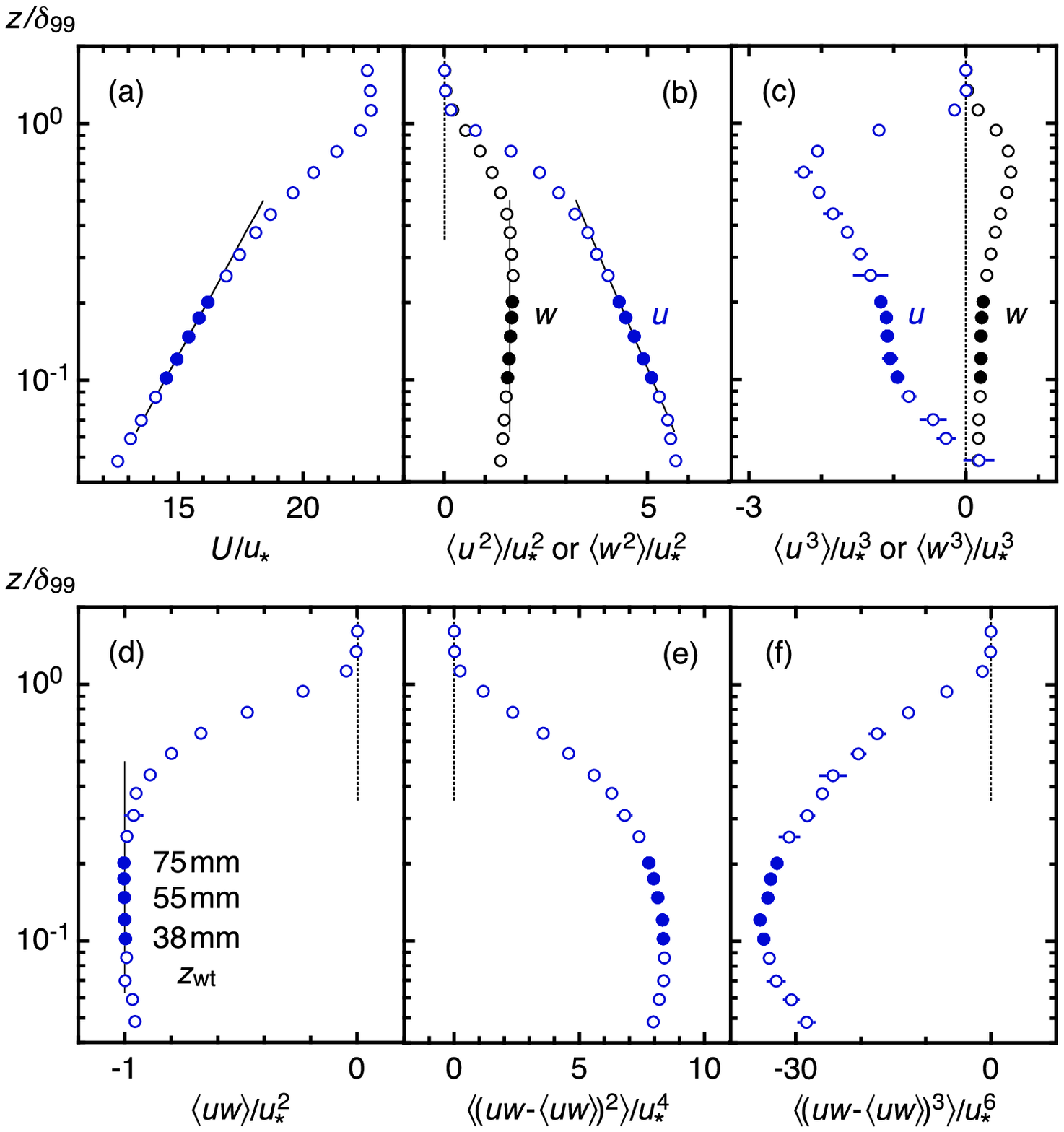}}
\caption{\label{f2} One-time statistics for a range of $z /\delta_{99}$: (a) $U/u_{\ast}$, (b) $\langle u^2 \rangle/u_{\ast}^2$ or $\langle w^2 \rangle/u_{\ast}^2$, (c) $\langle u^3 \rangle/u_{\ast}^3$ or $\langle w^3 \rangle/u_{\ast}^3$, (d) $\langle uw \rangle/u_{\ast}^2$, (e) $\langle (uw - \langle uw \rangle)^2 \rangle/u_{\ast}^4$, and (f) $\langle (uw - \langle uw \rangle)^3 \rangle/u_{\ast}^6$. The filled symbols lie within the constant-flux sublayer. To these, solid lines are least-squares fits of Eq.~(\ref{eq1b}) or (\ref{eq10}). We provide $\pm 2\sigma$ errors, albeit not including those for $u_{\ast}$ and $\delta_{99}$.}
\end{center}
\end{figure} 

\subsection{Overall characteristics of data}

Figure~\ref{f2} shows one-time statistics semi-logarithmically as a function of $z/\delta_{99}$. To exemplify the measurement uncertainties, we provide $\pm 2\sigma$ errors on all the data, albeit usually not discernible.

At least from $z_{\rm wt} = 38$ to $75$\,mm or from $z/\delta_{99} = 0.10$ to $0.20$ (filled circles), the mean momentum flux $\langle uw \rangle$ remains a constant. Its value is used to estimate the friction velocity as $u_{\ast} = (-\langle uw \rangle)^{1/2}$, which is in turn used to estimate the parameters of laws of the mean velocity $U$ in Eq.~(\ref{eq1b}) and so on (solid lines). The results are summarized in Table \ref{t1}. Within $\pm 2\sigma$ errors, they are consistent with values in the literature (see also Sec. \ref{S5b}).

The inner bound of our constant-flux sublayer has been affected by the size and separation of roughness rods on the wall surface.\cite{fs14} We observed effects of the individual rods up to $z_{\rm wt} \simeq 30$\,mm, coinciding with $z/\delta_{99} \simeq 0.08$, by shifting measurement positions slightly in the $x_{\rm wt}$ direction.

The fluctuations of $u$, $w$, and $uw$ in units of $u_{\ast}$ or $u_{\ast}^2$ are large. As shown in Figs.~\ref{f2}(c) and (f), they are skewed positively for $w$ or negatively for $u$ and $uw$.\cite{fsc07, smhffhs18}

\begin{figure}[tbp]
\begin{center}
\resizebox{8.6cm}{!}{\includegraphics*[2.3cm,19.1cm][17.0cm,27.0cm]{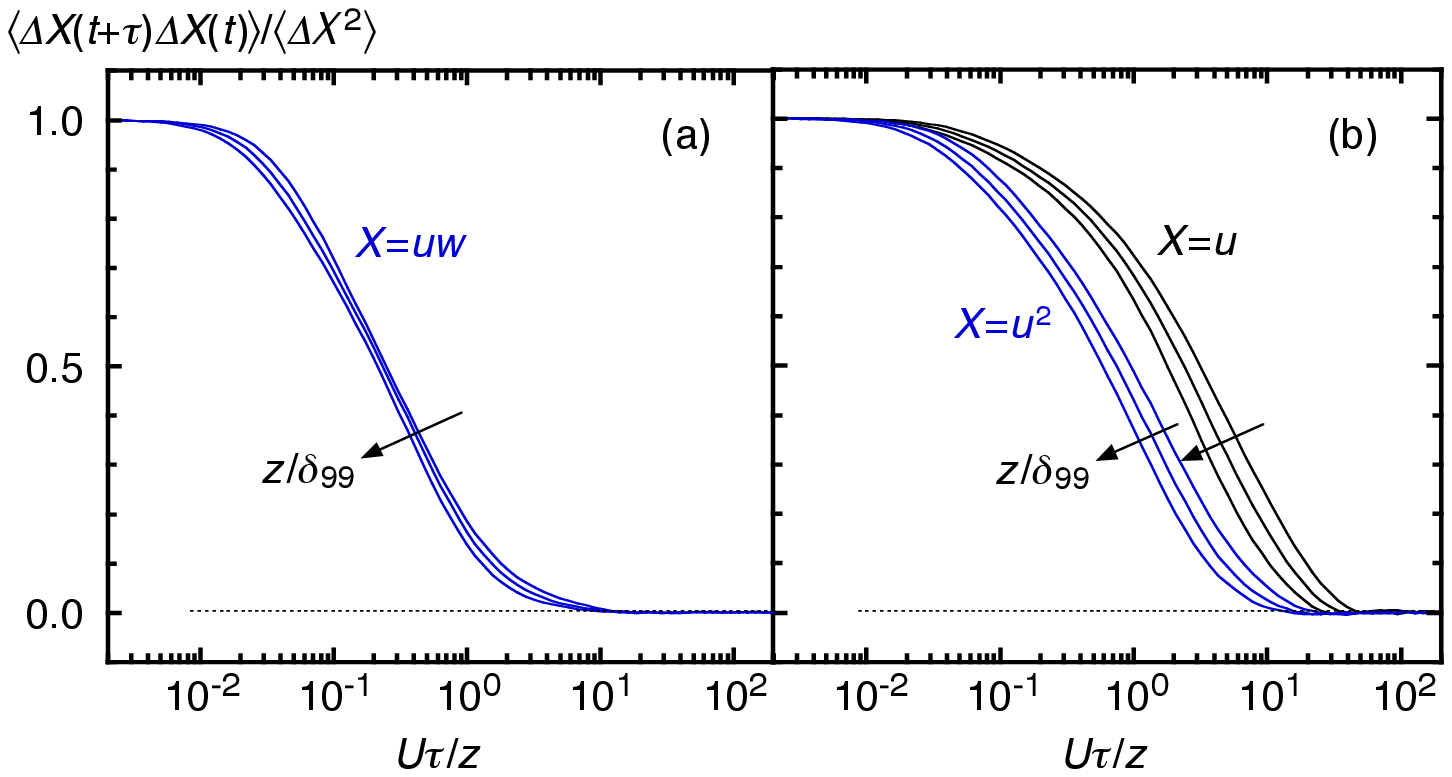}}
\caption{\label{f3} Two-time correlation $\langle {\mit\Delta}X(t+\tau){\mit\Delta}X(t) \rangle / \langle {\mit\Delta}X^2 \rangle$ at $z/\delta_{99} = 0.10$, $0.15$ and $0.20$ for a range of $U\tau/z$: (a) ${\mit\Delta}X = uw - \langle uw \rangle$ and (b) ${\mit\Delta}X = u$ or $u^2 - \langle u^2 \rangle$. The arrows indicate increasing $z/\delta_{99}$.}
\end{center}
\end{figure}

Figure~\ref{f3} shows two-time non-dimensional correlations $\langle {\mit\Delta}X(t+\tau){\mit\Delta}X(t) \rangle/\langle {\mit\Delta}X^2 \rangle$ for ${\mit\Delta}X = uw - \langle uw \rangle$ or $u^2 - \langle u^2 \rangle$ at $z_{\rm wt} = 38$, $55$, and $75$\,mm or $z/\delta_{99} = 0.10$, $0.15$ and $0.20$ in the constant-flux sublayer. They are persistent up to a timescale $\tau \simeq 10z/U$. While the correlation for $u^2 - \langle u^2 \rangle$ is dependent on $z/\delta_{99}$, such a dependence is not significant in that for $uw - \langle uw \rangle$.

The correlation for ${\mit\Delta}X = u$ in Fig.~\ref{f3}(b) is persistent up to a larger scale. If $u$ is approximated as a Gaussian process,\cite{my71, mouri17} since we have $\langle u_1 u_2 u_3 u_4 \rangle = \langle u_1 u_2 \rangle \langle u_3 u_4 \rangle + \langle u_1 u_3 \rangle \langle u_2 u_4 \rangle + \langle u_1 u_4 \rangle \langle u_2 u_3 \rangle$ with $u_n = u(t_n)$,
\begin{equation} \label{eq7}
\frac{\langle u^2(t+\tau) u^2(t) \rangle - \langle u^2 \rangle^2}{\langle u^4 \rangle - \langle u^2 \rangle^2} = \left[ \frac{\langle u(t+\tau)u(t) \rangle}{\langle u^2 \rangle} \right]^2 \le 1.
\end{equation}
Thus, the non-dimensional correlation for $u^2 - \langle u^2 \rangle$ decays to zero faster than that for $u$.

We also consider the Kolmogorov length $\eta$. If the mean energy dissipation $\langle \varepsilon \rangle$ is estimated as $15 \nu \langle (\partial u /U \partial t)^2 \rangle$, we have $\eta = \nu^{3/4}/ \langle \varepsilon \rangle^{1/4} = 0.16 \pm 0.01$\,mm at $z_{\rm wt} = 38$ to $75$\,mm in the constant-flux sublayer. Since the~length of our hot wire is $1.25$\,mm (Sec.~\ref{S3a}), small scales in the dissipation range might have been filtered out. Nevertheless, our focus is on the energy-containing scales, which have not been affected at all.

\begin{figure}[bp]
\begin{center}
\resizebox{8.6cm}{!}{\includegraphics*[2.3cm,3.7cm][17.0cm,27.0cm]{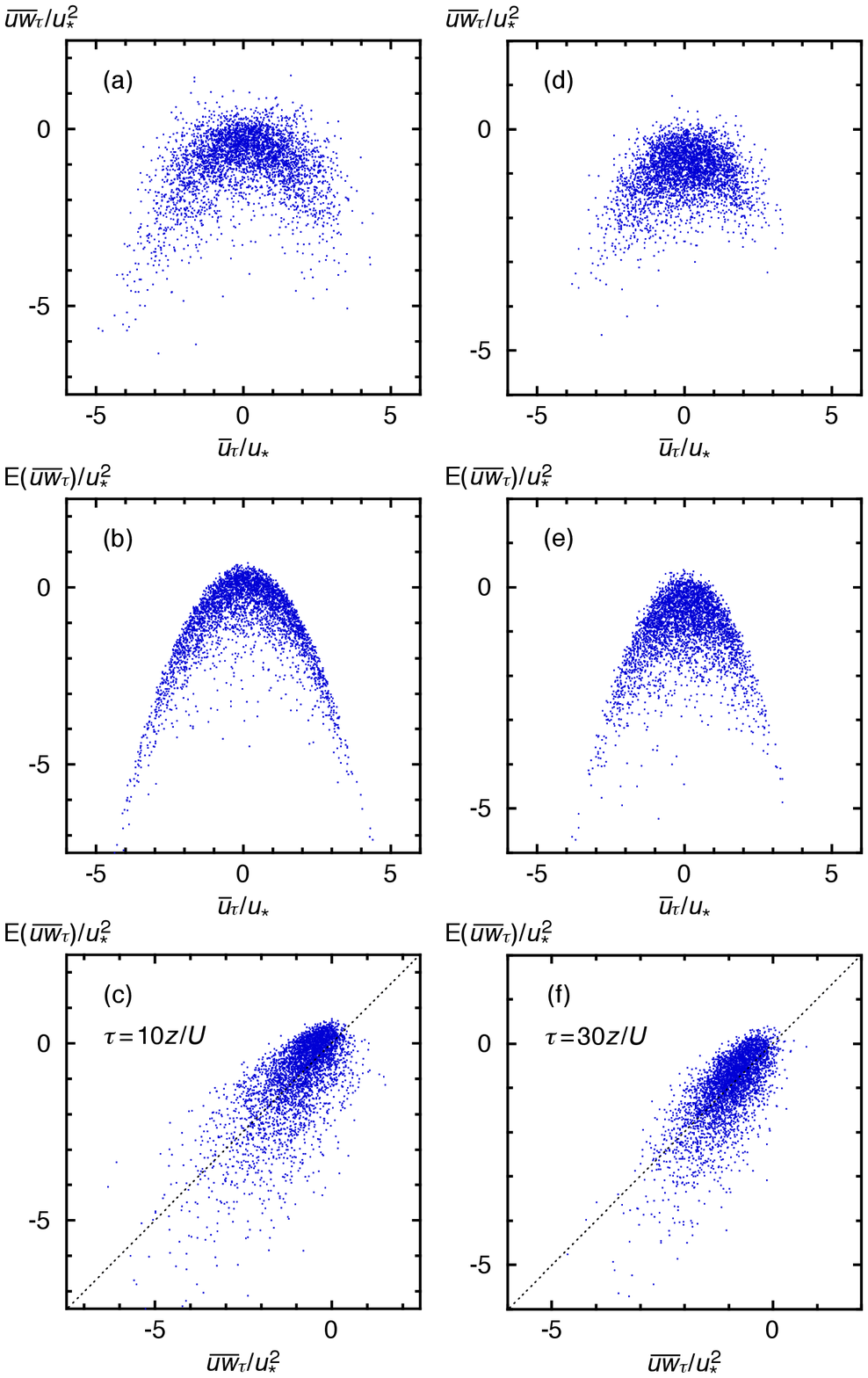}}
\caption{\label{f4} Scatter plots among $\overline{uw}_{\tau}/u_{\ast}^2$, ${\rm E}(\overline{uw}_{\tau})/u_{\ast}^2$, and $\overline{u}_{\tau}/u_{\ast}$ at $z/\delta_{99} = 0.15$ for $\tau = 10z/U$ (a--c) or $30z/U$ (d--f). Here ${\rm E}(\overline{uw}_{\tau})$ is an estimate of $\overline{uw}_{\tau}$ via Eq.~(\ref{eq3}).}
\end{center}
\end{figure} 
\begin{figure}[tbp]
\begin{center}
\resizebox{8.6cm}{!}{\includegraphics*[2.3cm,19.1cm][17.0cm,27.0cm]{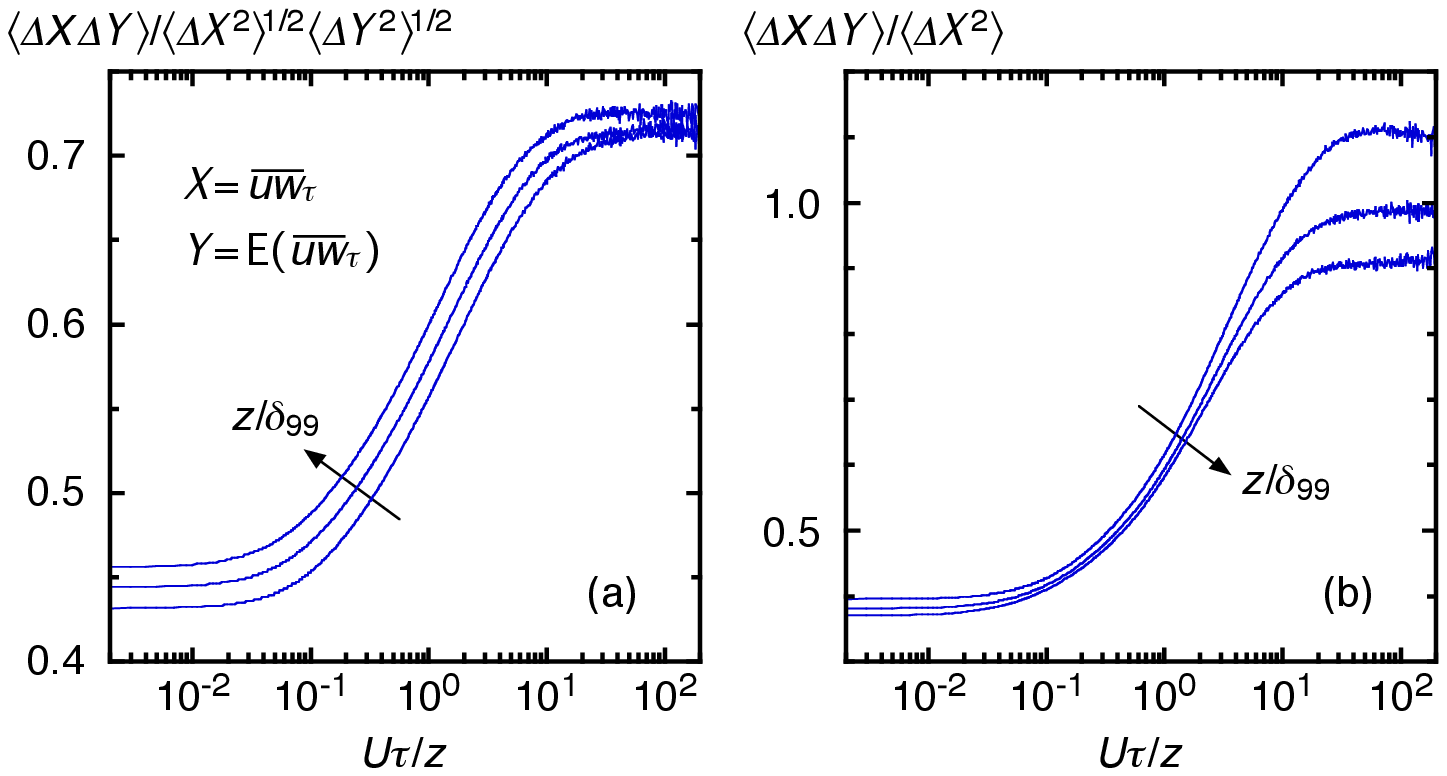}}
\caption{\label{f5} Correlation coefficient $\langle {\mit\Delta}X{\mit\Delta}Y \rangle / \langle {\mit\Delta}X^2 \rangle^{1/2} \langle {\mit\Delta}Y^2 \rangle^{1/2}$ (a) and correlation slope $\langle {\mit\Delta}X{\mit\Delta}Y \rangle / \langle {\mit\Delta}X^2 \rangle$ (b) between ${\mit\Delta}X = \overline{uw}_{\tau} - \langle \overline{uw}_{\tau} \rangle$ and ${\mit\Delta}Y = {\rm E}(\overline{uw}_{\tau}) - \langle {\rm E}(\overline{uw}_{\tau}) \rangle$ at $z/\delta_{99} = 0.10$, $0.15$, and $0.20$ for a range of $U\tau/z$. Here ${\rm E}(\overline{uw}_{\tau})$ is an estimate of $\overline{uw}_{\tau}$ via Eq.~(\ref{eq3}). The arrows indicate increasing $z/\delta_{99}$.}
\end{center}
\begin{center}
\resizebox{8.6cm}{!}{\includegraphics*[2.3cm,19.1cm][17.0cm,27.0cm]{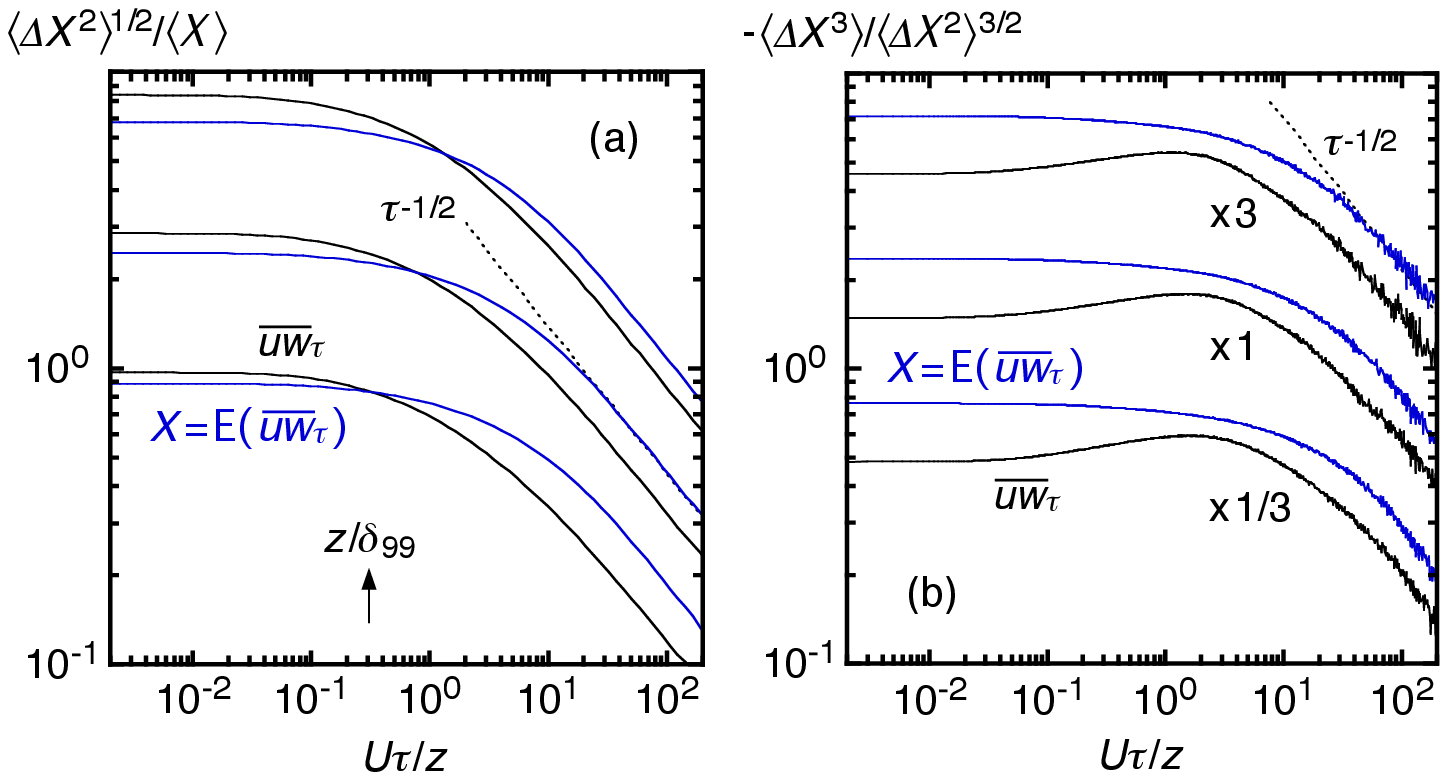}}
\caption{\label{f6} Standard deviation $\langle {\mit\Delta}X^2 \rangle^{1/2}/\langle X \rangle$ (a) and reversed-sign skewness $-\langle {\mit\Delta}X^3 \rangle/\langle {\mit\Delta}X^2 \rangle^{3/2}$ (b) for $X = \overline{uw}_{\tau}$ or ${\rm E}(\overline{uw}_{\tau})$ with ${\mit\Delta}X = X - \langle X \rangle$ at $z/\delta_{99} = 0.10$ (shifted by a factor $1/3$), $0.15$ (not shifted), and $0.20$ (shifted by a factor $3$) for a range of $U\tau/z$. Here ${\rm E}(\overline{uw}_{\tau})$ is an estimate of $\overline{uw}_{\tau}$ via Eq.~(\ref{eq3}).}
\end{center}
\end{figure} 

\section{Results} \label{S4}

To confirm Eq.~(\ref{eq3}), our data of Sec.~\ref{S3} at $z/\delta_{99} = 0.10$, $0.15$, and $0.20$ in the constant-flux sublayer are studied here. We use the measured values $\langle uw \rangle$, $\langle u^2 \rangle$, and $\overline{u^2}_{\tau}$ to obtain estimates of the momentum flux ${\rm E}(\overline{uw}_{\tau})$ via Eq. (\ref{eq3a}) and compare them with the measured values $\overline{uw}_{\tau}$ for a range of the smoothing timescale $\tau$.

Figure \ref{f4} shows scatter plots among $\overline{uw}_{\tau}$, ${\rm E}(\overline{uw}_{\tau})$, and the smoothed velocity $\overline{u}_{\tau}$ at $z/\delta_{99} = 0.15$ for the smoothing timescale $\tau = 10z/U$ or $30z/U$. Upon plots of $\overline{uw}_{\tau}$ against $\overline{u}_{\tau}$ in Figs.~\ref{f4}(a) and (d), we observe convex distributions such that the momentum flux $\overline{uw}_{\tau}$ is enhanced negatively when the streamwise velocity $U+\overline{u}_{\tau}$ deviates from its average $U$.\cite{im21} Those of ${\rm E}(\overline{uw}_{\tau})$ in Figs.~\ref{f4}(b) and (e) are similar. Actually in Figs.~\ref{f4}(c) and (f), data points are distributed along a dotted line of $\overline{uw}_{\tau} = {\rm E}(\overline{uw}_{\tau})$.

The negative deviation $\overline{u}_{\tau} < 0$ tends to be larger than the positive deviation $\overline{u}_{\tau} > 0$ in Figs.~\ref{f4}(a) and (d). Albeit not distinguished in Eq.~(\ref{eq3a}),  $u < 0$ with $w > 0$ is more important to $\langle uw \rangle < 0$ than $u > 0$ with $w < 0$ at $z/\delta \gtrsim 0.01$ so long as $z$ lies in the constant-flux sublayer.\cite{wbe72, wl72, lw73, fsc07, nkpk07, w16, dm21}

The results in Fig.~\ref{f4} appear to confirm our formula of Eq.~(\ref{eq3}). However, since the data points are yet scattered from the dotted line of $\overline{uw}_{\tau} = {\rm E}(\overline{uw}_{\tau})$ in Figs.~\ref{f4}(c) and (f) especially when the magnitude of $\overline{uw}_{\tau}$ is large, we study statistics of that scatter.

Figure \ref{f5}(a) shows the correlation coefficient between $\overline{uw}_{\tau}$ and ${\rm E}(\overline{uw}_{\tau})$ at $z/\delta_{99} = 0.10$, $0.15$, and $0.20$ against the smoothing timescale $\tau$ in units of $z/U(z)$. With an increase in $\tau$, the minor component $(u_{\perp}, w_{\perp})$ is smoothed away. The coefficient becomes large and then remains a constant $0.72 \pm 0.02$ at $\tau \gtrsim 10z/U$.

This constant for the correlation coefficient is less than unity. Being contrary to our expectation in Sec.~\ref{S2}, some fraction of the minor component $(u_{\perp}, w_{\perp})$ has too large timescales to be smoothed away. There might also exist an additional component (Sec.~\ref{S5b}). Such a result is not avoidable because we estimate $uw$ from $U+u$ alone. For example, in our preliminary estimation of $\overline{uw}_{\tau}$ based on machine learning of experimental time-series data,\cite{im21} the correlation coefficient had a similar value.

Figure \ref{f5}(b) shows the correlation slope of ${\rm E}(\overline{uw}_{\tau})$ on $\overline{uw}_{\tau}$ against the smoothing timescale $\tau$. With an increase in $\tau$, the slope becomes large and then remains a constant at $\tau \gtrsim 10z/U$. This constant is at $1.0 \pm 0.1$ as expected for $\overline{uw}_{\tau} \simeq {\rm E}(\overline{uw}_{\tau})$. It has yet a dependence on $z/\delta_{99}$, which is induced by a difference between those of ${\rm E}(\overline{uw}_{\tau})$ and $\overline{uw}_{\tau}$ (see also Sec.~\ref{S5a}).

Figure \ref{f6} shows the non-dimensional standard deviation $\langle {\mit\Delta}X^2 \rangle^{1/2} / \langle X \rangle$ and skewness $\langle {\mit\Delta}X^3 \rangle/\langle {\mit\Delta}X^2 \rangle^{3/2}$ for $X = \overline{uw}_{\tau}$ or ${\rm E}(\overline{uw}_{\tau})$ with ${\mit\Delta}X = X - \langle X \rangle$. Even at $\tau \simeq$ $10z/U$, the standard deviation is not so small, i.e., $30$ to $40$\% of its value at $\tau = 0$. The skewness is almost the same. As for $z/\delta_{99} = 0.15$ (middle pair of lines) at $\tau \gtrsim 10z/U$, the statistics of $\overline{uw}_{\tau}$ are $70$\% of those of ${\rm E}(\overline{uw}_{\tau})$.

Each statistic in Fig.~\ref{f6} follows a power law $\tau^{-1/2}$ at $\tau \gtrsim 10z/U$ (dotted line). Since the two-time correlation for ${\mit\Delta}uw$ is negligible there (Fig.~\ref{f3}), ${\mit\Delta}\overline{uw}_{\tau}(t)$ is an average of independent and identically distributed random variables $\sum_{n=1}^N {\mit\Delta}\overline{uw}_{\tau_0}(t_n) /N$ for $N = \tau/\tau_0$, $\tau_0 \simeq 10z/U$, and $t_n = t-\tau/2+(n-1/2)\tau_0$. We have $\langle {\mit\Delta}\overline{uw}_{\tau}^2 \rangle = \langle {\mit\Delta}\overline{uw}_{\tau_0}^2 \rangle /N$, $\langle {\mit\Delta}\overline{uw}_{\tau}^3 \rangle  = \langle {\mit\Delta}\overline{uw}_{\tau_0}^3 \rangle /N^2$, and those laws $\tau^{-1/2}$ as $N^{-1/2}$. The same applies to the case of ${\mit\Delta}\overline{u^2}_{\tau} \varpropto {\mit\Delta}{\rm E}(\overline{uw}_{\tau})$.\cite{mht09}

Thus, if the smoothing timescale is $\tau \gtrsim 10z/U$, our formula of Eq.~(\ref{eq3}) holds as a good approximation. Between the estimates ${\rm E}(\overline{uw}_{\tau})$ and the measured values $\overline{uw}_{\tau}$, the correlation coefficient is $0.7$. Its slope is about $1.0$. The standard deviation and skewness of ${\rm E}(\overline{uw}_{\tau})$ do not differ significantly from those of $\overline{uw}_{\tau}$.

\section{Discussion} \label{S5}

\subsection{Reconsideration of velocity decomposition} \label{S5a}

Our model for Eq.~(\ref{eq3a}) has used the reverse-sign von K\'arm\'an constant $-\kappa$ as the inclination $w_{\shortparallel}/u_{\shortparallel}$ of a major component of the velocity fluctuations (Sec.~\ref{S2}). We reconsider this inclination $w_{\shortparallel}/u_{\shortparallel}$ from a standard statistical approach, i.e., principal component analysis\cite{jc16} of the data.

The principal components are defined to be orthogonal eigenvectors of the covariance matrix at a given distance $z$,
\begin{subequations} \label{eq8}
\begin{equation} \label{eq8a}
\left[
\begin{array}{cc}
\langle u^2 \rangle & \langle uw  \rangle \\
\langle uw  \rangle & \langle w^2 \rangle
\end{array}
\right]
.
\end{equation}
Its eigenvalues are
\begin{equation} \label{eq8b}
\lambda = \frac{\langle u^2 \rangle + \langle w^2 \rangle \pm \sqrt{ (\langle u^2 \rangle - \langle w^2 \rangle )^2 + 4 \langle uw \rangle^2}}{2} .
\end{equation}
We thereby obtain the eigenvectors. Those for the larger and the smaller eigenvalues are assigned to the major and the minor components $(u_{\shortparallel}, w_{\shortparallel})$ and $(u_{\perp}, w_{\perp})$. Then,
\begin{equation} \label{eq8c}
\frac{w_{\shortparallel}}{u_{\shortparallel}} = \gamma - \sqrt{1+\gamma^2} < 0
\ \mbox{and} \
\frac{w_{\perp}}{u_{\perp}} = \gamma + \sqrt{1+\gamma^2} > 0,
\end{equation}
with
\begin{equation} \label{eq8d}
\gamma = - \frac{\langle u^2 \rangle - \langle w^2 \rangle}{2 \langle uw \rangle} > 0
\ \mbox{for} \
\langle u^2 \rangle > \langle w^2 \rangle
\ \mbox{and} \
\langle uw \rangle < 0.
\end{equation}
\end{subequations}
It is just like fitting an ellipse to data points in Fig.~\ref{f1}(a), by ignoring $\langle u^3 \rangle < 0$ and $\langle w^3 \rangle > 0$ in Fig.~\ref{f2}(c). If the $u$-$w$ plane is rotated so as to maximize the variance along one of its axes, the direction of such an axis is identical to that of Eq.~(\ref{eq8c}) for the major component $w_{\shortparallel}/u_{\shortparallel}$.

\begin{figure}[tbp]
\begin{center}
\resizebox{8.6cm}{!}{\includegraphics*[2.3cm,19.1cm][17.0cm,27.0cm]{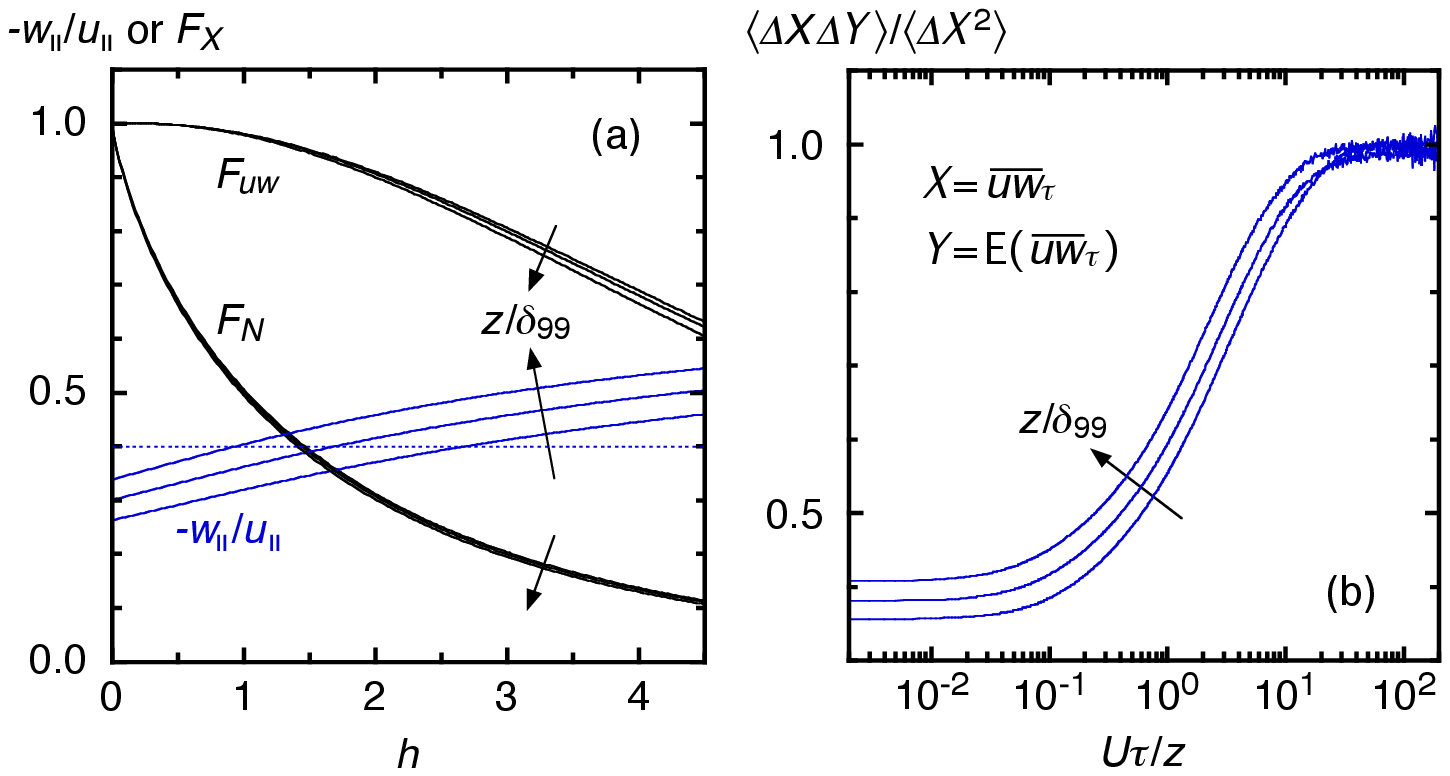}}
\caption{\label{f7} (a) Reversed-sign inclination $-w_{\shortparallel}/u_{\shortparallel}$ in Eq.~(\ref{eq8c}) and fraction $F_N$ as well as fractional contribution $F_{uw}$ in Eq.~(\ref{eq9b}) for $| uw | > h \langle uw \rangle$ in Eq.~(\ref{eq9a}) at $z/\delta_{99} = 0.10$, $0.15$, and $0.20$. The arrows indicate increasing $z/\delta_{99}$. (b) Same as in Fig.~\ref{f5}(b) but obtained by replacing $\kappa = 0.40$ in Eq.~(\ref{eq3a}) with $0.36$ and $0.44$ respectively for $z/\delta_{99} = 0.10$ and $0.20$.}
\end{center}
\end{figure}

We also follow Willmarth and Lu\cite{wl72, lw73} to impose a threshold $h$ on the magnitude of $uw$ as
\begin{subequations}
\begin{equation} \label{eq9a}
\vert uw \vert > h \langle uw \rangle .
\end{equation}
If this condition is satisfied by instantaneous values of $uw$ at $t_1$, $t_2$, ..., and $t_{N(h)}$ in a time series that has the total number $N_T$, their fraction $F_N(h)$ and their fractional contribution to the mean momentum flux $F_{uw}(h)$ are
\begin{equation} \label{eq9b}
F_N(h) = \frac{N(h)}{N_T} 
\ \mbox{and} \
F_{uw}(h) = \frac{1}{N(h)} \sum_{n=1}^{N(h)} \frac{uw(t_n)}{\langle uw \rangle} .
\end{equation}
\end{subequations}
For such data, we calculate the inclination $w_{\shortparallel}/u_{\shortparallel}$ of Eq. (\ref{eq8c}) as a function of the threshold $h$. The results are summarized in Fig.~\ref{f7}(a).

The inclination $w_{\shortparallel}/u_{\shortparallel}$ depends both on the threshold $h$ and on the wall-normal distance $z$. With an increase in $h$ for each of $z$, it changes across $w_{\shortparallel}/u_{\shortparallel} = -0.40$ (dotted line), i.e., the value of $-\kappa$ in Eq.~(\ref{eq3a}). As for $z/\delta_{99} = 0.15$ (middle line), $w_{\shortparallel}/u_{\shortparallel}$ is equal to $-0.40$ at $h \simeq 2$. There is a systematic difference by $\pm 10$\% for $z/\delta_{99} = 0.10$ and $0.20$.

This difference explains the dependence of the correlation slope on $z/\delta_{99}$ in Fig.~\ref{f5}(b). By replacing $\kappa = 0.40$ in Eq.~(\ref{eq3a}) with other constants $0.36$ and $0.44$ respectively for $z/\delta_{99} = 0.10$ and $0.20$, we correct our estimates of the momentum flux ${\rm E}(\overline{uw}_{\tau})$. The resultant slopes in Fig.~\ref{f7}(b) are all equal to unity at $\tau \gtrsim 10z/U$. Since the correlation lies essentially between ${\mit\Delta}\overline{uw}_{\tau}$ and ${\mit\Delta}\overline{u^2}_{\tau} \varpropto$ ${\mit\Delta}{\rm E}(\overline{uw}_{\tau})$, its coefficient remains the same.

On the other hand, as observed in Fig.~\ref{f7}(a), the fraction $F_N$ and the fractional contribution $F_{uw}$ are dependent only on the threshold $h$. For the above case $h \simeq 2$, while $F_N$ has decayed significantly, $F_{uw}$ is still close to unity.\cite{wl72, lw73, w16, nkpk07, dm21} That is, fluctuations at $h \lesssim 2$ are not important to the mean momentum flux $\langle uw \rangle$. They have cancelled out one another. The values of $w_{\shortparallel}/u_{\shortparallel}$ at $h \lesssim 2$ imply that $w$ tends to be weak with respect to $u$ in those fluctuations.\cite{wbe72}

To conclude, $w_{\shortparallel}/u_{\shortparallel} = -\kappa$ used for Eq.~(\ref{eq3a}) represents fluctuations that dominate the momentum flux $uw$. This holds especially at around the middle of the constant-flux sublayer such as $z/\delta_{99} \simeq 0.15$ in our experiment. At the other distances $z$, we might need to correct for some deviation of the inclination $w_{\shortparallel}/u_{\shortparallel}$ from the value of the reverse-sign von K\'arm\'an constant $-\kappa$.

\subsection{Implication from phenomenology of eddies} \label{S5b}

Wall turbulence is often modeled as a random superposition of eddies that are attached to the wall.\cite{t76, mouri17, mm19} They have the same shape but are of various sizes. The larger eddies are increasingly rare. If the wall-normal size of such an eddy is much larger than the observing distance $z$, it contributes only to the wall-parallel velocities $U+u$ and $v$. If that size is less than the distance $z$, there is no contribution. For the constant-flux sublayer, Townsend\cite{t76} has derived laws of velocity variances,
\begin{equation} \label{eq10}
\frac{\langle u^2(z) \rangle}{u_{\ast}^2} = c_{u^2} + d_{u^2} \ln \left( \frac{\delta}{z} \right)
\ \mbox{and} \
\frac{\langle w^2(z) \rangle}{u_{\ast}^2} = c_{w^2} .
\end{equation}
While the constants $c_{u^2}$ and $c_{w^2}$ originate in eddies with wall-normal sizes comparable to the observing distance $z$, the logarithmic term $d_{u^2} \ln (\delta /z)$ originates in eddies with sizes from $z$ to the turbulence thickness $\delta$. Boundary layers are known to have $c_{u^2} \simeq 2.0$--$2.5$, $d_{u^2} \simeq 1.2$--$1.3$,\cite{mmhs13, mmym17, smhffhs18} and $c_{w^2} \simeq 1.3$--$1.6$.\cite{amt00, fsc07, fs14, mkbm15, mmym17} These values are consistent with ours in Table \ref{t1}.

The streamwise size of an attached eddy has been assumed to be about $10$ times its wall-normal size.\cite{bhm17, mm19} Actually at $\tau \gtrsim 10z/U$, the correlation for $uw$ in Fig.~\ref{f3}(a) is negligible. Upon smoothing over such a timescale, most of the minor component $(u_{\perp}, w_{\perp})$ is smoothed away (Fig. \ref{f5}). We attribute this to internal fluctuations of the eddies. The rest of the minor component and also the major component $(u_{\shortparallel}, w_{\shortparallel})$ are attributable to net contributions of the individual eddies.

Since the attached eddies are allowed to overlap freely with one another, they need not be identical to motions organized in actual wall turbulence,\cite{mm19} e.g., packets of hairpin-shaped vortices.\cite{amt00, dn11a} Nevertheless, $uw$ is known to be enhanced by these motions.\cite{amt00, dn11b, w16}

The attached eddies with wall-normal sizes from $z$ to $\delta$ are not important to $w$ and hence explain the behavior of $w_{\shortparallel}/u_{\shortparallel}$ in Fig.~\ref{f7}(a). With an increase in $z/\delta$ or $h$, these eddies become rare. We have $\langle u^2 \rangle /u_{\ast}^2 \rightarrow c_{u^2}$ in Eq.~(\ref{eq10}) and $\gamma \rightarrow (c_{u^2}-c_{w^2})/2$ in Eq.~(\ref{eq8d}). The values of $c_{u^2}$ and $c_{w^2}$ in Table \ref{t1} lead to $-w_{\shortparallel}/u_{\shortparallel} \rightarrow 0.67$ in Eq.~(\ref{eq8c}), being consistent with the observed behavior.

The contributions from those large eddies might be regarded as a component that is independent of $(u_{\shortparallel}, w_{\shortparallel})$ and $(u_{\perp}, w_{\perp})$. However, since their timescales are large, their contributions are not removed by smoothing in Eq. (\ref{eq3b}) alone. We have instead incorporated them into the inclination of the major component $w_{\shortparallel}/u_{\shortparallel} = -\kappa = -0.40$ for Eq.~(\ref{eq3a}) as discussed in Sec.~\ref{S5a}.

Finally, we remark on very large-scale structures of actual wall turbulence, i.e., streamwise alignments of aforementioned organized motions. They are meandering with lengths $\gtrsim 20\delta$.\cite{ka99, dn11b} Although they might modulate $uw$ over a scale $\varpropto \delta$,\cite{mmch13, lkbb16, dm21} such a modulation is not important at least to our modeling. From power laws $\tau^{-1/2}$ at $\tau \gtrsim 10 z/U$ for the standard deviation and skewness of $\overline{uw}_{\tau}$ (Fig.~\ref{f6}), it follows that $\delta$ does not affect such statistics. To these, fluctuations at the smaller scales are rather dominant.

\subsection{Extension to thermally stratified cases} \label{S5c}

Thus far, we have focused on thermally neutral cases. If the wall surface is horizontal and is heated or cooled with respect to the overlying flow, it is unstable or stable as is usual in the atmosphere. The constant-flux sublayer is still existent, but we need to correct its law of Eq.~(\ref{eq1a}).  According to Monin and Obukhov,\cite{my71,f06}
\begin{subequations}
\begin{equation} \label{eq11a}
\frac{z}{u_{\ast}} \frac{dU}{dz} = \frac{\phi_U (z/L_{\rm O})}{\kappa} 
\ \ \mbox{with} \ \
\phi_U(0) = 1.
\end{equation}
Here, a non-dimensional function $\phi_U$ is to be determined experimentally or observationally. The Obukhov length $L_{\rm O}$ is dependent not only on the mean momentum flux $\langle uw \rangle$ but also on the mean heat flux: $L_{\rm O} < 0$ in unstable cases, $L_{\rm O} > 0$ in stable cases, and $L_{\rm O} \rightarrow \pm\infty$ in the neutral limit. We could rely on this theory except under highly stable conditions.\cite{f06, m14}

By using Eq.~(\ref{eq11a}) instead of Eq.~(\ref{eq1a}) in our model of Sec.~\ref{S2}, we replace the inclination of the major component $w_{\shortparallel}/u_{\shortparallel} = -\kappa$ with $w_{\shortparallel}/u_{\shortparallel} = -\kappa/\phi_U$ to reformulate the momentum flux $uw$ as
\begin{equation} \label{eq11b}
\overline{uw}_{\tau}(z,t) - \langle uw \rangle = -\frac{\kappa}{\phi_U (z/L_{\rm O})} \left[ \overline{u^2}_{\tau}(z,t) - \langle u^2(z) \rangle \right] .
\end{equation}
\end{subequations}
This is a relation between $\overline{uw}_{\tau} - \langle uw \rangle$ and $\overline{u^2}_{\tau} - \langle u^2 \rangle$ as in the original formula of Eq.~(\ref{eq3a}). For the smoothing, its timescale $\tau$ is to be determined as a function of $z$, $U$, and also $L_{\rm O}$.

Monin and Obukhov\cite{my71,f06} also considered the temperature and the concentration of a passive scalar. The wall-normal flux of such a quantity $X$ is $w (X - \langle X \rangle) = w {\mit\Delta}X$. Its average $\langle w {\mit\Delta}X \rangle$ is constant in the constant-flux sublayer. Being analogous to Eq.~(\ref{eq11a}),
\begin{subequations}
\begin{equation} \label{eq12a}
\frac{z}{u_{\ast}} \frac{d \langle X \rangle}{dz} = \frac{\langle w {\mit\Delta}X \rangle}{\langle uw \rangle} \frac{\phi_{\langle X \rangle} (z/L_{\rm O})}{\kappa} .
\end{equation}
Here $\phi_{\langle X \rangle}(0)$ is not necessarily equal to unity. By comparing Eq.~(\ref{eq12a}) with Eq.~(\ref{eq11a}), we generalize Eq.~(\ref{eq11b}) as
\begin{align} \label{eq12b}
&\overline{w{\mit\Delta}X}_{\tau}(z,t) - \langle w{\mit\Delta}X \rangle = \\
&\quad                                 - \frac{\langle uw \rangle}{\langle w {\mit\Delta}X \rangle}
                                         \frac{\kappa}{\phi_{\langle X \rangle} (z/L_{\rm O})} \left[ \overline{{\mit\Delta}X^2}_{\tau}(z,t) - \langle {\mit\Delta}X^2(z) \rangle \right] . \nonumber
\end{align}
\end{subequations}
The smoothing timescale $\tau$ is likely the same as that for Eq.~(\ref{eq11b}) because the same eddies transfer all of the quantities. For example, albeit under a thermally neutral condition, a logarithmic law like Eq.~(\ref{eq10}) is observed for the variance of a scalar concentration.\cite{mmym17}

The formula of Eq.~(\ref{eq3a}) is thus extended not only to the momentum flux $uw$ in thermally stratified cases but also to other fluxes there. It is desirable to confirm Eqs. (\ref{eq11b}) and (\ref{eq12b}) in laboratory experiments or field observations.

\section{Concluding Remarks} \label{S6}

For the constant-flux sublayer of wall turbulence, the law of the wall of Eq.~(\ref{eq1}) relates the mean momentum flux $\langle uw \rangle$ to the mean streamwise velocity $U$ at the same wall-normal distance $z$. However, the corresponding relation is still uncertain between instantaneous values of the momentum flux $uw$ and the streamwise velocity $U+u$. We have explored such a relation. The momentum transfer is dominated by the velocity component $u_{\shortparallel}$ with $w_{\shortparallel} = -\kappa u_{\shortparallel}$ (Fig.~\ref{f1}). Via temporal smoothing to single out this component, we have obtained Eq.~(\ref{eq3}).

To confirm Eq.~(\ref{eq3}), we have studied experimental time-series data of a boundary layer (Fig.~\ref{f2} and Table~\ref{t1}). They are consistent with Eq.~(\ref{eq3}) if the smoothing timescale is $\tau \gtrsim 10z/U$ (Figs.~\ref{f4}--\ref{f6}), at which the two-time correlation for $uw$ is negligible (Fig.~\ref{f3}).

Since the constant-flux sublayer is essentially the same, Eq.~(\ref{eq3}) with $\tau \gtrsim 10z/U$ applies not only to boundary layers but also to pipe flows, channel flows, and so on. They would nevertheless exhibit differences, e.g., as for the coefficient and slope of the correlation between the actual flux and its estimate such as in Fig.~\ref{f5}. A significant difference is in fact known for the parameter $c_{u^2}$ of Eq.~(\ref{eq10}). It is at $2.0$--$2.5$ in boundary layers\cite{mmhs13, mmym17, smhffhs18} but is at $1.4$--$1.9$ in pipe flows if we use the pipe radius as the thickness $\delta$.\cite{mmhs13, ofsbta17}

The derivation of Eq.~(\ref{eq3}) has assumed strict stationarity of the turbulence. It still applies to a non-stationary flow if the mean streamwise direction, the mean streamwise velocity $U$, and the mean momentum flux $\langle uw \rangle$ are well defined over some averaging timescale that exceeds the smoothing timescale of $\tau \gtrsim 10 z/U$.

We also remark on an application to a large-eddy simulation, i.e., a numerical simulation used widely for wall turbulence. It is to resolve all energy-containing eddies, but such an eddy is too small in the vicinity of the wall surface. The first off-wall grid points are usually in the constant-flux sublayer, for which we need a model to estimate instantaneous values of the momentum flux $uw$ and those of the stress at the wall surface.\cite{bp18, lkbb16}

To estimate the momentum flux $uw$ from the streamwise velocity $U+u$, available even at the first off-wall grid points, Eq.~(\ref{eq2}) is often used as the simplest model.\cite{d70, bmp05, kl12, lkbb16, ypm17, bp18, bl21, im21} However, the resultant profile of the mean velocity $U$ is known to differ from the law of Eq.~(\ref{eq1b}). This difference depends largely on details of the simulation, e.g., grid configuration and subgrid-scale modeling,\cite{kl12, lkbb16, bp18, bl21} but Eq. (\ref{eq2}) itself is not consistent with fluctuations of $uw$ as in Fig.~\ref{f1}(b).\cite{ypm17, bp18, im21}

The formula of Eq.~(\ref{eq3}) serves as an alternative model. Being in contrast to Eq.~(\ref{eq2}), it reproduces negative enhancements of the momentum flux $uw$ against deviations of the streamwise velocity $U+u$ from its average $U$ (Fig. \ref{f4}). There is also an extension to thermally stratified cases in the form of Eq.~(\ref{eq11b}) or (\ref{eq12b}). A numerical study for applying these formulae to a large-eddy simulation of wall turbulence is highly desirable.

\begin{acknowledgments}
This work was supported in part by KAKENHI Grant No. 19K03967. We are grateful to T. Yagi and K. Mori for their help during the experiment.
\end{acknowledgments}

\end{document}